\documentclass[journal=ancac3,manuscript=article]{achemso}

\usepackage[utf8]{inputenc}
\usepackage{textcomp}  
\usepackage{gensymb}  
\usepackage{graphicx}  
\usepackage{amsmath} 
\usepackage{amssymb} 
\usepackage{color}

\author{Robert J. S. Ivancic}
\affiliation[University of Pennsylvania]
  {Department of Physics and Astronomy, University of Pennsylvania, Philadelphia, PA 19104}
\author{Robert A. Riggleman}
\email{rrig@seas.upenn.edu}
\affiliation[University of Pennsylvania]
  {Department of Chemical and Biomolecular Engineering, University of Pennsylvania, Philadelphia, PA 19104}

\title[Shear banding in polymer nanopillars]
  {Identifying structural signatures of shear banding in polymer nanopillars}


\begin{document}

\begin{tocentry}
 \includegraphics[width=9cm]{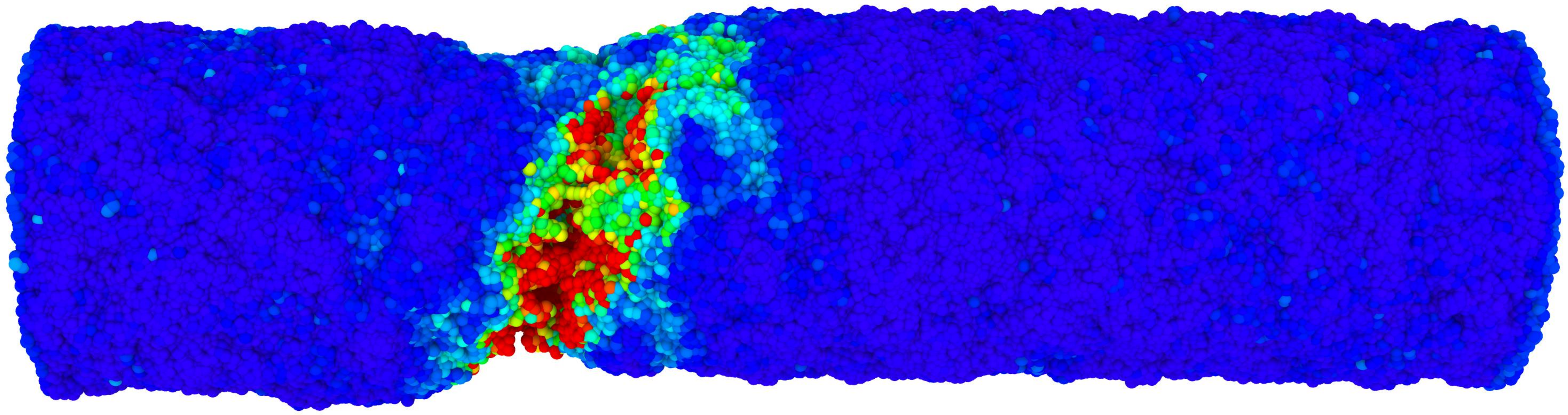}
\label{fig:TOC}
\end{tocentry}

\begin{abstract}
Amorphous solids are critical in the design and production of nanoscale devices, but under strong confinement these materials exhibit changes in their mechanical properties which are not well understood. Phenomenological models explain these properties by postulating an underlying defect structure in these materials but do not detail the microscopic properties of these defects. Using machine learning methods, we identify mesoscale defects that lead to shear banding in polymer nanopillars well below the glass transition temperature as a function of pillar diameter. Our results show that the primary structural features responsible for shear banding on this scale are fluctuations in the diameter of the pillar. Surprisingly, these fluctuations are quite small compared to the diameter of the pillar, less than half of a particle diameter in size. At intermediate pillar diameters, we find that these fluctuations tend to concentrate along the minor axis of shear band planes. We also see the importance of mean “softness” as a classifier of shear banding grow as a function of pillar diameter. Softness is a new field that characterizes local structure and is highly correlated with particle-level dynamics such that softer particles are more likely to rearrange. This demonstrates that softness, a quantity that relates particle-level structure to dynamics on short time and length scales, can predict large time and length scale phenomena related to material failure.
\end{abstract}

There are numerous applications where amorphous organic materials are used in highly confined geometries, including as polymer photoresists in semiconductor manufacturing\cite{stoykovich_directed_2005}, the active layers in organic light-emitting diodes\cite{zhang_efficient_2014, kim_multicolored_2015}, and in polymer nanocomposites at high loadings of nanoparticles\cite{huang_polymer_2015, hor_nanoporous_2017}. In many of these applications, in particular semiconductor manufacturing, the mechanical properties of the confined material are of utmost importance. Generally speaking, amorphous materials have many unique mechanical properties including high strength, high stiffness, and low mechanical dissipation\cite{ashby_engineering_2012, jones_engineering_2012, trexler_mechanical_2010, jaeger_granular_1996, bansal_handbook_2013, parmenter_mechanical_1998, landel_mechanical_1993}. These properties make them desirable in a number of engineering applications; however, their use is hindered by their tendency to fail in a brittle manner\cite{robertson_diamond-like_2002, ashby_metallic_2006, herrmann_nanoparticle_2007, zhang_using_2013, hiemenz_principles_1997}. A hallmark of these catastrophic failure modes is shear banding, the localization of shear strain to a narrow region which develops during deformation \cite{bigoni_nonlinear_2012, manning_strain_2007}. Shear banding has been experimentally observed in many types of amorphous materials including: granular materials \cite{fenistein_kinematics:_2003, tsai_granular_2005}, bubble rafts \cite{lauridsen_shear-induced_2002, kabla_local_2003}, complex fluids \cite{mair_observation_1996, makhloufi_rheo-optical_1995}, and metallic glasses \cite{lu_deformation_2003, johnson_deformation_2002}.

Although shear banding has been extensively studied in the bulk using phenomenological models, a microscopic theory of shear banding has proven elusive. The phenomenological models that describe shear banding can broadly be classified into two types. Solid mechanics models postulate some constitutive relations about how a material behaves at each point in space. In these theories, a shear band forms when a small region of the material has a perturbed set of constitutive relations causing it to shear more easily\cite{rice_initiation_1973, rice_localization_1976, bordignon_strain_2015}. Similarly, mean-field models, including shear transformation zones\cite{falk_dynamics_1998, langer_shear-transformation-zone_2008}, soft glassy rheology\cite{sollich_rheology_1997}, and others\cite{zheng_mesoscopic_2009}, hypothesize mesocale ``configurational soft spots"\cite{manning_strain_2007}, regions that are more likely to yield under shear stress, and these regions propagate to form a shear band. While these two types of theories have significantly different starting points, they both predict that shear bands form from mesoscale defects in a solid but provide few details as to the nature of these defects. Although some indirect estimates of their volume are available \cite{pan_correlation_2009, choi_estimation_2012}, the microscopic structure that underlies these defects is unknown \cite{schuh_mechanical_2007}. Moreover, it is unclear whether bulk defects are the primary cause of shear banding in confined materials. Previous work has shown that the location of strain localization is somehow quenched into the molecular structure when forming a glass\cite{shavit_strain_2014}, suggesting that the local structure could play a key role.

In this study, we examine a large set of molecular dynamics simulations of amorphous oligomeric nanopillars that are strained to failure.  Using a novel machine learning method, we detect mesoscale structural defects which lead to shear band formation. We systematically vary the pillar diameter in these systems from $12.5$ -- $100$ monomer diameters to understand how these defects vary as the system becomes less confined and more bulk-like. From this defect structure, we make quantitative predictions about where shear bands will form. Our machine learning approach allows us to look at a broad array of structural features and perform an unbiased selection of those which correlate with shear banding at each pillar diameter. Here, we pay special attention to another machine-learned microscopic structural quantity, ``softness," which is strongly predictive of particle-level rearrangements in disordered materials \cite{cubuk_identifying_2015}. Softer particles have structures which make them more likely to rearrange than harder (less soft) particles. This quantity has been implicated in the understanding of aging glasses \cite{schoenholz_relationship_2017} and the universal yield strain in bulk disordered materials \cite{cubuk_structure-property_2017}, but the connection between softness and mesoscale phenomena such as shear banding has not been explored.

We find that small fluctuations in the diameter of the pillar, less than $\frac{1}{2}$ of a particle diameter in size, are most predictive of where shear bands will form in these pillars regardless of the diameter of the pillar. This is surprising as these surface fluctuations are not mechanically induced (from dust for example) but come about from the thermalization of the pillars themselves. We also find that our coarse grained softness features become more important for distinguishing whether a plane will shear band as pillar diameter increases. Planes that are softer than average are more likely to shear band. To ensure the density features are not sufficient to predict shear banding alone, we verify that these softness features do better than random chance at identifying shear bands even in the absence of correlations with other density features. 

The importance of these results is twofold. First, they suggest that small surface defects induced during the thermalization of nanoscale amorphous components may play a major role in their mechanical properties up to the micron scale. Indeed, these results suggest that focusing on manufacturing processes that lead to smooth surfaces as opposed to hard interiors will yield stronger nanoscale materials. Second, more fundamentally, they suggest that softness may be the microscopic origin of mesoscale configurational soft spots in the bulk. This connection is non-trivial as we are relating a structural quantity (that is associated with local, short-time scale dynamics) to shear band formation, a non-local, long-time scale event. Even more interesting, we find that we do not need to know the dynamical nature of these defects as we approach the shear banding event. Knowing their configuration prior to deformation is sufficient. This suggests that at temperatures well below the glass transition temperature these defects are locked in place.

\section{Results}

Our polymer model is a modified coarse-grained oligomer with five Lennard-Jones interaction sites per chain, and the monomers of each chain are connected with stiff harmonic bonds. The Lennard-Jones potential used in this work is modified to promote shear banding and fracture at temperatures far below the glass transition $T_g$. We prepare nanoscale cylinders by equilibrating our system at temperatures above $T_g$ in a simulation box that is periodic along the length before slowly quenching to $T = 0.05$, which is far below our simulated glass transition temperature $T_g = 0.38$. We note that all quantities are reported in reduced Lennard-Jones units, and complete details of the model are provided in the methods below. 

\begin{figure}
\centering
\includegraphics[width=4in]{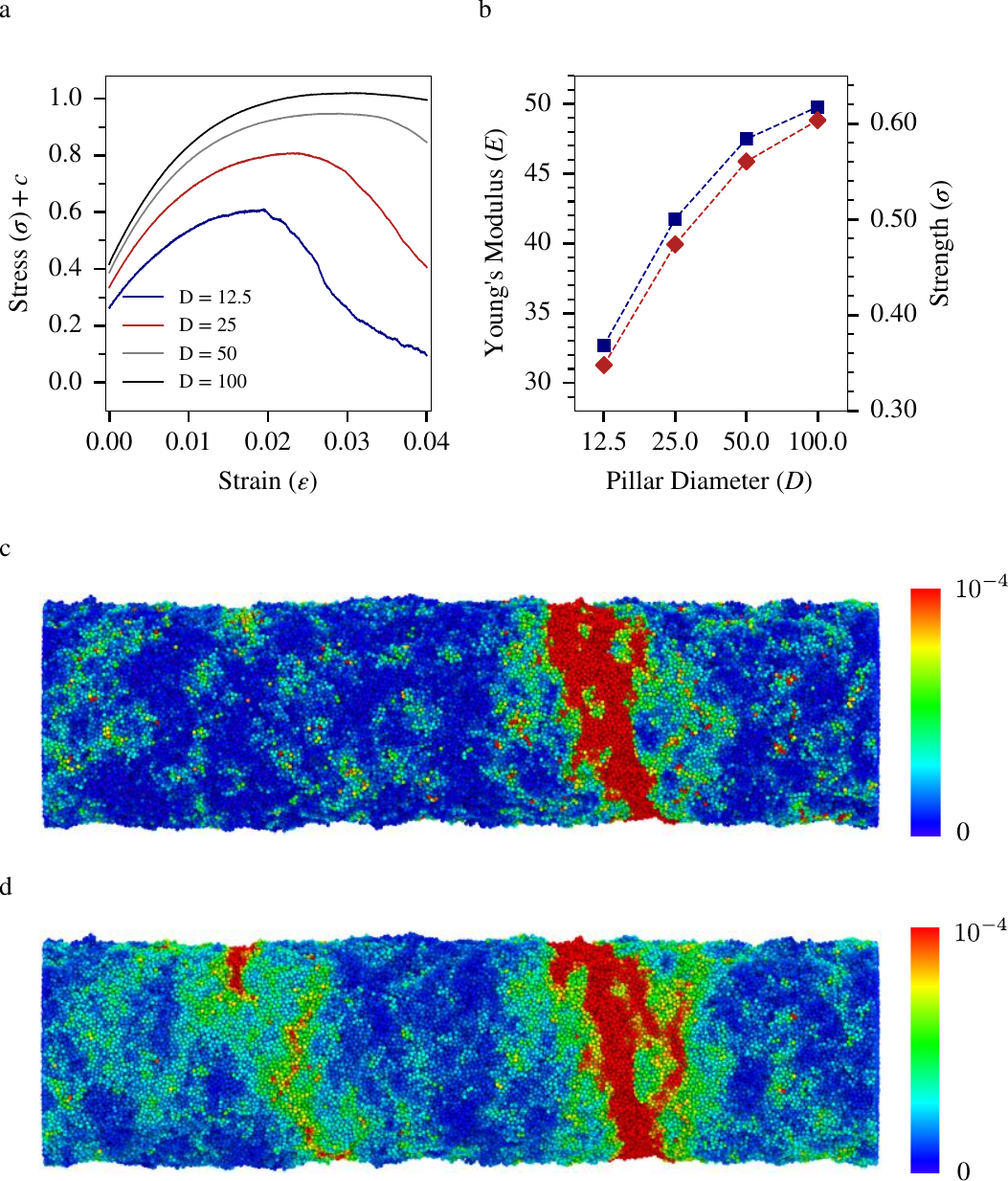}
\caption{
\textbf{Characterization of basic mechanical properties of oligomer nanopillars}
\textbf{(a)} Stress-strain curves averaged over all configurations found for each nanopillar diameter when deformed under uxiaxial tension at an engineering strain rate $\dot \epsilon = 2.5 \times 10^{-5}$ at $T = 0.05$. The curves are vertically shifted for clarity.
\textbf{(b)} Young's modulus (navy squares) and the strength (red diamonds) of the nanopillars as a function of the pillar diameter.
\textbf{(c)} The von Mises strain field of a single $D = 50$ pillar calculated by comparing the rearrangements surrounding each particle after a strain of $\epsilon = 5.5\%$, and
\textbf{(d)} the von Mises strain field averaged over $50$ $D = 50$ pillars in the isoconfigurational ensemble. 
}
\label{fig:MechanicalProperties}
\end{figure}

Figure \ref{fig:MechanicalProperties} shows that the mechanical properties of our pillars depends strongly on the pillar diameter. To deform our samples, we applied a uniaxial strain to the $\hat{z}$ axis at an engineering strain rate of $\dot{\epsilon} = 2.5 \times 10^{-5}$ at $T = 0.05$. We plot engineering stress-strain curves averaged over all configurations at each pillar diameter in Figure \ref{fig:MechanicalProperties}a. We find that both the Young's modulus, which was determined by linear fits to the initial ($\epsilon \leq 0.005$) stress-strain response, and the strength (stress maximum) of our pillars increases with pillar diameter. Both material properties increase by more than $50$ percent as the pillar diameter increases from $D=12.5$ to $D=100$ as shown in Figure \ref{fig:MechanicalProperties}b. The overall trends with sample dimension are in good qualitative agreement with experiments on thin polymer films as a function of film thickness \cite{stafford_buckling-based_2004, liu_directly_2015}.

The strain in our samples strongly localizes into a shear band as our deformations reach the yield point. To understand how deformation effects the strain field within our pillars, we examine the von Mises shear strain rate around each particle, denoted as $J_{2}$ which is a common metric in numerical studies of shear banding \cite{adibi_surface_2015, li_deformation-driven_2015, shimizu_theory_2007}. Figure~ \ref{fig:MechanicalProperties}c shows the von Mises strain rate field of a single $D = 50$ pillar, and this field exhibits an unambiguous shear band plane of high von Mises shear strain rate. At this low temperature, all of our samples at any pillar diameter exhibit a strong strain localization.

A key point we wish to address with our study is whether the location where a material fails is dictated by the local structure, and if so, we further wish to identify the structural motifs that promote strain localization and shear banding. To first test whether the local structure plays a role in the localization of a shear band, we employ the isoconfigurational ensemble \cite{widmer-cooper_central_2009}, which is a technique that played a key role in demonstrating that there exists an interplay between local structure and dynamic heterogeneities in supercooled liquids. By beginning a series of simulations with the same monomer positions, but with momenta re-drawn from the Maxwell-Boltzmann distribution, we can examine whether the location of the shear band in our pillar is caused by random thermal fluctuations or the material structure. If we begin with the same configuration used to generate the strain field in Figure~\ref{fig:MechanicalProperties}c and run 50 deformation trajectories with randomly initialized momenta, the average strain field $\langle J_{2,j}\rangle$ field for each particle $j$ is shown in Figure~\ref{fig:MechanicalProperties}d. Clearly the strain tends to localize in one of two locations, while if the location of the shear band were random, we would expect a more uniform distribution. These results indicate that the local structure that is frozen when the sample is quenched plays an important role in determining the shear band location, consistent with prior work \citenum{shavit_strain_2014}. Furthermore, this tendency for strain to localize is robust across all studied pillar diameters.

Having established that the local structure dictates where shear bands will form using the isoconfigurational ensemble, in order to guide the development of mesoscale and constitutive models it is essential to determine the nature of the structural variables that lead to strain localization. As a result, our next goal is to identify which structural motifs (e.g., the local density in the center of the pillar, or perhaps the local roughness on the surface) are associated with shear band formation. We approach this problem as one of classification in which we want to distinguish between two sets of planes: those that are likely to shear band and those that are not; these sets will be called ``weak" and ``strong" planes respectively. Thus, we aim to create an independent function for each pillar diameter, called a ``classifier", that can classify a plane into the weak or strong category at each pillar diameter based on its structure alone. Using specific classifiers for each pillar diameter allows for the possibility that the features which determine shear banding vary with pillar diameter.  To develop our classifier, we build a ``training set" of planes: one population that does shear band, and a second population that does not shear band, which are defined based on the largest and smallest average von Mises shear strain rate in a pillar, respectively. These planes are selected from a set of $50$ or more independent pillar thermalizations and deformations at each pillar diameter.

To solve this classification problem, each candidate plane is characterized by $M$ ``structure functions", which encode the density and local softness distribution as a function of radial position in the plane, distance away from the plane, or angular slices through the major and minor plane axes. Each plane $i$ is assigned a vector $\boldsymbol{p}_i$ with $M$ elements that each correspond to a distinct structure function.  A linear support vector machine (SVM) finds the best hyperplane to separate shear band and non-shear band structure vectors in $\mathbb{R}^{M}$. We define the ``weakness" of a plane $i$,  $W_{i}$, to be the shortest signed distance from $\boldsymbol{p}_{i}$ to this hyperplane in $\mathbb{R}^{M}$. Larger values of plane weakness indicate planes that are structurally similar to shear banding planes while smaller values of $W_i$ indicate little structural similarity to shear banding planes. This hyperplane is then employed to determine the plane weakness of any plane at a given pillar diameter. We normalize our hyperplane so that the distribution of plane weakness has a standard deviation of $1$. Our SVM method was implemented using scikit-learn \cite{pedregosa_scikit-learn:_2011}, and recursive feature elimination allows us to ensure that our models are not overfit\cite{guyon_gene_2002}.

\begin{figure}
\centering
\includegraphics[width=4.5in]{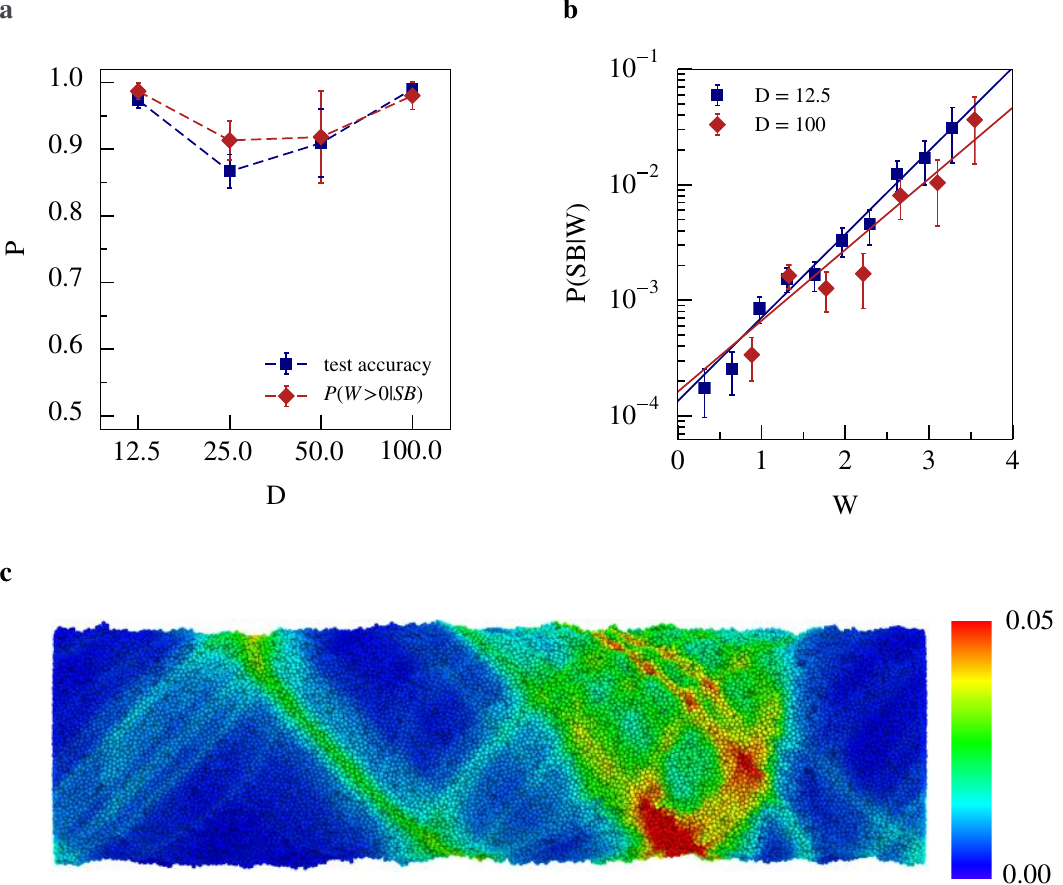}
\caption{
\textbf{Performance of plane weakness as structural indicator of shear banding planes.}
\textbf{(a)} Test set accuracy (navy squares) and expected percentage of shear bands that are weak (red diamonds) at all pillar diameters.
\textbf{(b)} The probability that a plane will shear band as a function of its weakness at pillar diameters $D = 12.5$ and $D = 100$. Solid lines are exponential fits to the data.
\textbf{(c)} A snapshot of an undeformed $D = 50$ pillar where each monomer $j$ is colored by  $P_j$. Error bars in the above fits are calculated using a binomial confidence interval.
}
\label{fig:Performance}
\end{figure}

Figure \ref{fig:Performance}a demonstrates that our classifiers are able to distinguish shear banding planes from non-shear banding planes at each pillar diameter. The test set accuracy gives an unbiased estimate of the percentage of shear band and non-shear band planes that are correctly classified. At each pillar diameter over $85\%$ of planes are correctly classified, which is $8$ standard errors above random ($50\%$) proving that we do better than chance at distinguishing between shear band and non-shear band planes. The second metric, $P\left(W>0|\text{SB}\right)$, provides the probability that a shear band plane ($\text{SB}$) is classified as weak ($W>0$). We find that over $90\%$ of shear band planes are weak at each pillar diameter. These results show that our linear SVMs correctly classify the vast majority of shear band planes as weak.

Now we consider the predictive nature of plane weakness' magnitude rather than its sign alone. We plot the probability a plane will shear band for a given plane weakness, $P\left(\text{SB}|W\right)$, in Figure \ref{fig:Performance}b for the $D = 12.5$ and $D = 100$ pillars. We see an exponential increase by more than $2$ decades over the range $W = 0$ to $W = 3$ in the probability of shear banding, and the trends are remarkably similar across pillar diameter, despite the fact that each diameter is characterized by a distinct classifier. This plot explicitly demonstrates that the probability of a shear banding is a function of magnitude, not just the sign, of plane weakness. As a plane becomes weaker as quantified by the local structure through $W_i$, it is more likely to shear band.

We next investigate whether there are spatial correlations in plane weakness that lead to regions in our sample that are more (or less) likely to shear band. To do so, we begin with $P(SB | W_i)$, the probability that plane $i$ of given weakness will shear band, and map it to the particles near the plane to estimate the probability that particle $j$ will be in a shear band,
\begin{equation}
P_{j} = 
\frac{\sum_{i} P\left(\text{SB}|W_{i}\right) \Theta_{ij}^{P}\left( 0, \xi_{h} \right)}
{\sum_{i}\Theta_{ij}^{P}\left( 0, \xi_{h} \right)}.
\label{eq:7}
\end{equation}
Here, the sum is over all planes, $\Theta_{ij}^{P} \left(h,  \xi_{h} \right) = e^{-\left(\lvert h_{ij} \rvert-h \right)^{2}/\xi_{h}^{2}}$ is a weighting function that controls the spatial extent of the mapping from plane $i$ to particle $j$, $h_{ij}$ is the distance between plane $i$ and particle $j$ and $\xi_h = 1/2$ is a parameter that controls the decay length of $\Theta_{ij}^P$. The map of $P_j$ for all particles is shown for a $D = 50$ pillar in Figure \ref{fig:Performance}c, and this is the same pillar configuration shown in Figures \ref{fig:MechanicalProperties}c and \ref{fig:MechanicalProperties}d. Evidently, spatial correlations exist in plane weakness leading to two large defect regions in the pillar where the particles are more likely to be involved in a shear band. The locations of high average von Mises shear strain rate seen in Figure \ref{fig:MechanicalProperties}d show striking similarities with regions of high $P_j$ in Figure \ref{fig:Performance}c. The Pearson correlation between these two plots is $0.52$, and the probability that there is no correlation between these fields is less than $10^{-6}$. This strong correlation demonstrates that plane weakness predicts not only the planes that are likely to fail but also the spatial regions that are likely to fail in a pillar. This distinction is important as it indicates that plane weakness is a \textit{direct} structural measure of these regions as opposed to an \textit{indirect} quantity that is only useful in plane space. We emphasize that what makes this result remarkable is that we are predicting the location of shear bands, a strongly nonlinear phenomenon, from the initial configuration prior to any deformation and then finding these results directly compare to the actual locations of failure.

Taken together the results in Figure \ref{fig:Performance} demonstrate the structural origin of shear banding in glassy polymer nanopillars. This leads to the question: which plane structures cause shear banding? Since the plane weakness $W_i$ is defined as the signed normal distance to a hyperplane in a space defined by our structure functions, a natural approach to determining the importance of various structure functions would be to consider the magnitude of the projection of the hyperplane normal onto each structure function axis. This approach, however, would assume that each structure function is independent and would not account for correlations in the structural information encoded between structure functions. In other words, it would assume our structure functions form an orthogonal basis in the high-dimensional space, which is clearly false in our case. As a result of multicollinearity and our fitting proceedure, slight differences in sampled data may lead to large differences in the perceived importance of various structure functions.

Instead, we will say that a structure function is important if varying that structure function is likely to cause a large variance in plane weakness. A metric for this is called the Feature Importance Ranking Measure (FIRM) \cite{zien_feature_2009}. A structure function’s FIRM score is the percentage of the variance in plane weakness that can be described by the variance in that structure function if correlations with other structure functions are included. As such, FIRM scores range between $0$, where the variance in plane weakness is not described by a given structure function, and $1$, where the variance of plane weakness is entirely described by variance of a given structure function. In the event that our structure functions are uncorrelated, FIRM simplifies to the projection of the structure function onto the hyperplane normal. 

\begin{figure}
	\centering
	\includegraphics[width=5.836in]{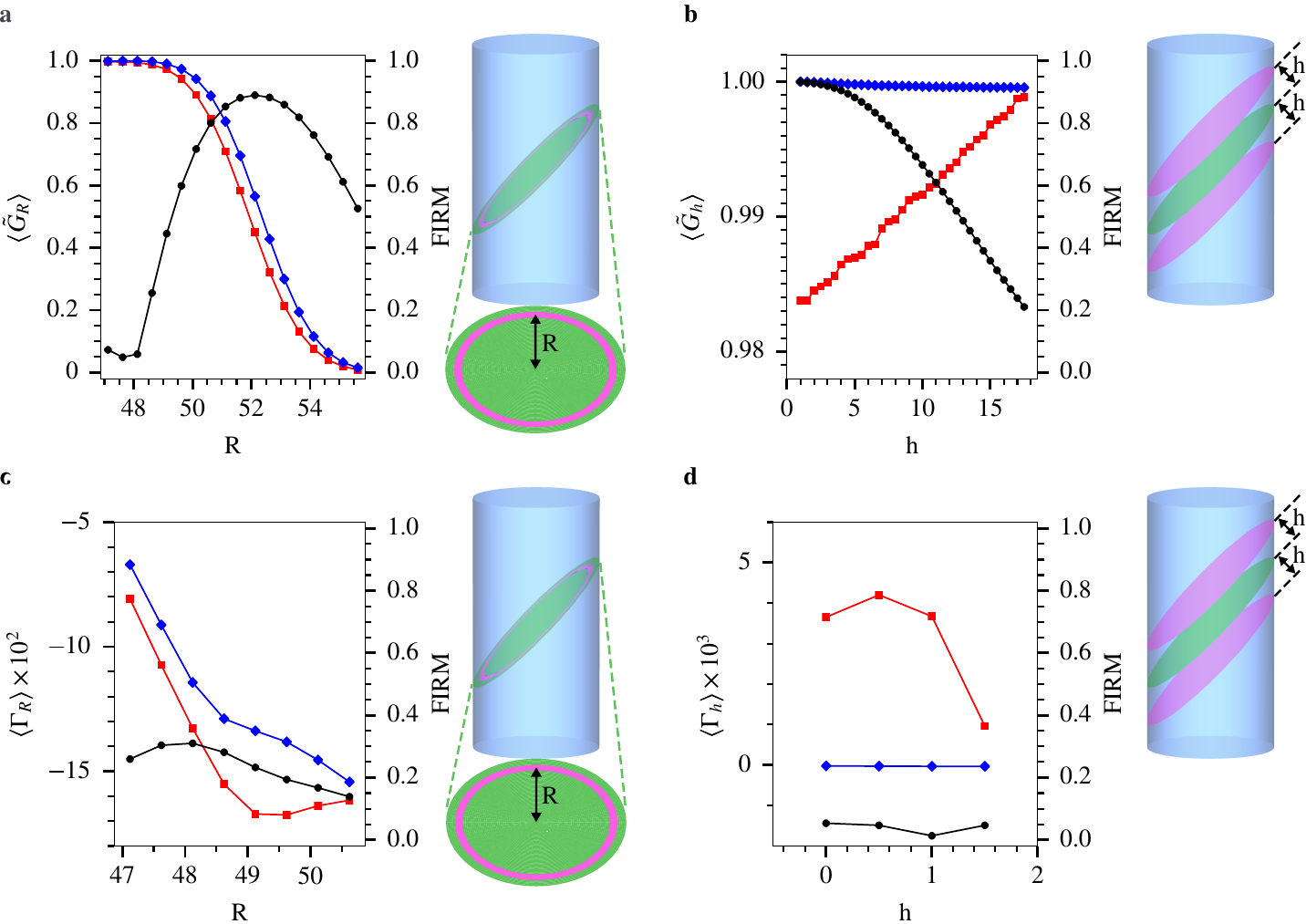}
	\caption{
		\textbf{Plots of structure functions averaged over all (blue diamonds) and weak (red squares) planes with corresponding FIRM scores (black circles).} The left hand axis corresponds to the average of the set of structure functions. The right hand axis corresponds to the FIRM score of the given structure function. The graphics depicted to the right of the plots illustrate the region over which each structure function is calculated. The green plane represents the plane of consideration while the magenta regions represent the region over which the density function is calculated. All functions are plotted for the $D = 100$ pillar.
		The functions these plots show are: 
		\textbf{(a)} $\langle \tilde{G}_{R} \left( i; 3.00, 0.5, R \right) \rangle$,
		\textbf{(b)} $\langle \tilde{G}_{h} \left( i; 0.5, h \right) \rangle$, 
		\textbf{(c)} $\langle \Gamma_{R} \left( i; 3.00, 0.5, R \right) \rangle$ and
		\textbf{(d)} $\langle \Gamma_{h} \left( i; 0.5, h \right)  \rangle$ for $h \leq 1.5$.
		Here, a tilde above the function indicates that it has been normalized by the maximum of the given structure function set averaged over all planes.
	}
	\label{fig:FIRM}
\end{figure}

Figure \ref{fig:FIRM} plots several of the structure functions along with their FIRM scores to demonstrate the relative importance of different structural variations to shear banding for pillars with $D = 100$. The structure function characterizing the density as a function of radial position in a given plane is shown in Figure \ref{fig:FIRM}a for shear-banding and all planes, where each point in the curve corresponds to a different structure function. In general, we see that average radius of a shear banding plane is slightly smaller than the average plane. What is surprising about this feature is how small the fluctuation in the radius is, less than $\frac{1}{2}$ of a particle diameter. This length scale is nearly constant at all pillar diameters (See supporting information). The FIRM score for the density variations is also the highest near the surface, indicating that the variations in the density near the cylinder surface can be used to explain a large fraction of the variations in the plane weakness. In contrast, the density further away from the interface (where $R \approx 48$) is a less important indicator, as shown by the FIRM scores that decrease below 0.1 for $R \lesssim 48$. Remarkably, these fluctuations are not due to any mechanical scraping of the surface of the pillars but arise from the thermal fluctuations in the formation of our pillars alone.

The remaining panels in Figure \ref{fig:FIRM} show the importance of some other families of structure functions that we have employed in our machine learning approach. Figure \ref{fig:FIRM}b shows the importance of the total density in a plane a distance $h$ away from the test plane. Intuitively, this function is very important for small $h$ (FIRM score above 0.8) where it characterizes the density close to the plane, and this function becomes decreasingly important as $h$ increases. This provides further confirmation of our previous results revealing the most important feature is a slight undercoordination of the shear band plane due to these small surface fluctuations. We also see that these surface defects are quite long ranged along the surface of the pillar, approximately 18 particle diameters for the $D = 100$ pillar. The length scale of these surface defects grows sub-linearly with pillar diameter, which suggests that surface defects may become less important as the pillar diameter increases. This is in qualitative agreement with capillary-wave model (CWM) theory for planar liquid-vapor interfaces which suggests that this length scale should increase with the system's interfacial area as these fluctuations can better explore large wavelength modes  \cite{bedeaux_correlation_1985} (See supporting information). This suggests that these surface fluctuations are trapped during the quench of our pillars.

As described above, the softness of a particle has been shown to be intimately related to the tendency for an individual particle to rearrange under mechanical deformation or thermal relaxation \cite{cubuk_structure-property_2017, schoenholz_structural_2016, schoenholz_relationship_2017, sussman_disconnecting_2017}. A natural question to ask is whether the softness of the particles associated with a given plane is in any way indicative of the tendency of that plane to shear band and lead to failure. In Figure \ref{fig:FIRM}c, we plot the structure functions characterizing the average softness as a function of radial position in the pillars. The shear banding planes tend to have smaller values of softness near their surface compared to average planes, suggesting that shear band planes are harder near the surface. Now, we plot the structure functions that describe the average softness as a function of distance away from a test plane, $h$, in Figure \ref{fig:FIRM}d. We note that shear band planes have larger values of softness for small $h$ than non-shear band planes. However, given the relatively small FIRM score for each of these softness-based structure functions, we find that softness is not as predictive of the structural variations in shear banding planes, and the other structure functions, such as the radial density shown in Figure \ref{fig:FIRM}a, are better able to distinguish shear-banding planes.

The results described above in Figure \ref{fig:FIRM} suggest that different families of structure functions can have varying amounts of overall importance, and a natural question to ask is how the importance of groups of structure functions might change with pillar diameter. However, the FIRM score in its current implementation is restricted to single structure function characterizations\cite{zien_feature_2009}; so to address this short-coming, in this work we extend FIRM to analyze the importance of multiple structure functions simultaneously. Our approach, the Multiple Feature Importance Ranking Measure (MFIRM), describes the percentage of the variance in plane weakness that can be ascribed to the variance in a given set of structure functions if we take correlations into account, and we use this metric to distinguish the importance of families of structure functions (e.g., surface density fluctuations, angular density fluctuations, etc.). MFIRM then enables us to examine how the importance of families of structure functions changes with pillar diameter and assess whether we approach a limit where the bulk-response dominates the behavior.

\begin{figure}
    \centering
    \includegraphics[width=2in]{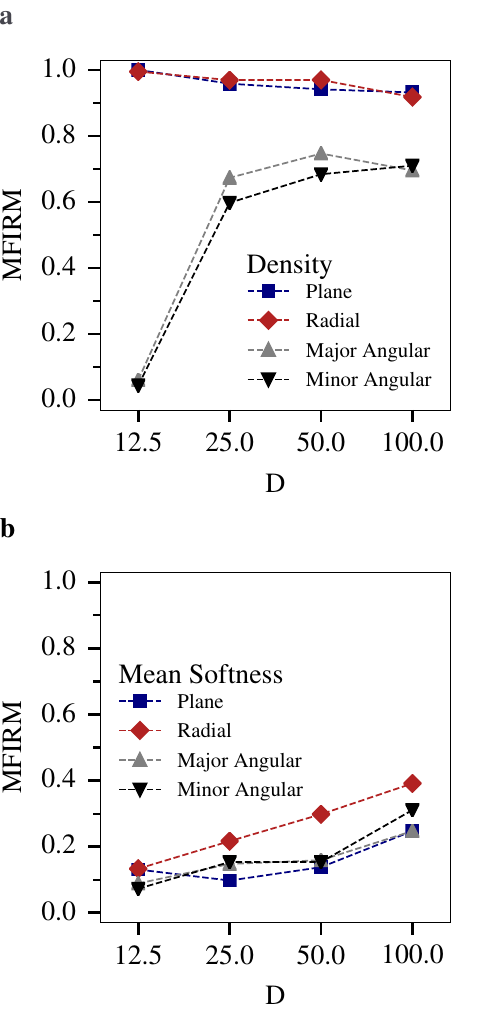}
    \caption{
        \textbf{Importance of sets of structure functions in shear band prediction.}  Plots of the MFIRM scores the plane, radial, and angular structure functions along the major and minor axes of each plane weighted by \textbf{(a)} the local density and \textbf{(b)} the mean softness as a function of pillar diameter $D$. These plots explain the percentage of the variance in plane weakness explained by each of these sets of features respectively.
    }
    \label{fig:MFIRM}
\end{figure}

Figure \ref{fig:MFIRM}a considers the MFIRM score of each family of functions weighted by the density at each pillar diameter $D$. The most striking feature of this plot is the large MFIRM scores of the radial and plane density structure functions which correspond to the sets of structure functions plotted in Figures \ref{fig:FIRM}a and \ref{fig:FIRM}b respectively. These structure functions account for more than $90$ percent of the variance in plane weakness at all pillar diameters. We note that it is possible to have multiple feature sets with high scores due to the correlation between the families of structure functions, an issue we account for below. The second important feature of Figure \ref{fig:MFIRM}a is the increasing MFIRM scores for angular density structure functions, which examine the density in angular slices along the minor and major axes of the ellipsoidal plane, with increasing pillar diameter. These scores explain around $70$ percent of the variance in plane weakness by $D = 25$, however these structure functions are unimportant for our smallest nanopillar. The MFIRM scores of the families of softness-based structure functions at each pillar diameter are shown in Figure \ref{fig:MFIRM}b. These softness-based structure functions measure mean softness in the same regions defined by the corresponding density structure functions above. Interestingly, the percentage of the variance in plane weakness these structure functions can explain increases with the pillar diameter, suggesting that softness functions become increasingly important as $D$ increases. We observe the two largest increases in MFIRM occur in the radial and minor angular mean softness structure functions. These sets of functions increase from accounting for $13$ and $7$ percent of the variance in plane weakness at $D = 12.5$ to $39$ and $31$ percent of the variance in plane weakness respectively.

The correlation (multi-collinearity) between structure functions makes it difficult to disentangle whether these high MFIRM scores represent a single underlying important variable (the radial fluctuations in the plane) or if the large MFIRM scores are a result of many such important variables. To ascertain which scenario is at play, we adopt the following approach. First, we hypothesize a set of structure functions that we believe may represent an underlying variable other than the radial fluctuations in the plane. Then, we fit this set of structure functions to the radial and plane density structure functions for all planes at a given pillar diameter using least squares multiple linear regression. We interpret this fit as a function that provides the expected value of the set of structure functions given a plane's radial and plane density structure functions which clearly measure these radial fluctuations in the plane. We next calculate the residuals between the actual and expected structure function values. We call these residuals the ``fluctuations" away from the structure function set's expected value. We then train a new machine learning hyperplane based exclusively on these fluctuations to obtain plane weakness, thus creating a metric that distinguishes between shear band and non-shear band planes based exclusively on these fluctuations. If a set of structure functions contains latent variables that are not described by the radial and plane density structure function model, then the fluctuations captured by these structure functions should be predictive of shear banding. Because much of the strength of plane weakness is attributable to these radial fluctuations (large MFIRM scores), we do not expect these models to be especially predictive. However, we may conclude that the more predictive these fluctuations are the greater the strength of the underlying latent variables that are not degenerate with the radial and plane density structure functions alone. In general, we denote these models based on fluctuations away from the radial and plane density functions as ``fluctuation models".

\begin{figure}
    \centering
    \includegraphics[width=5.836in]{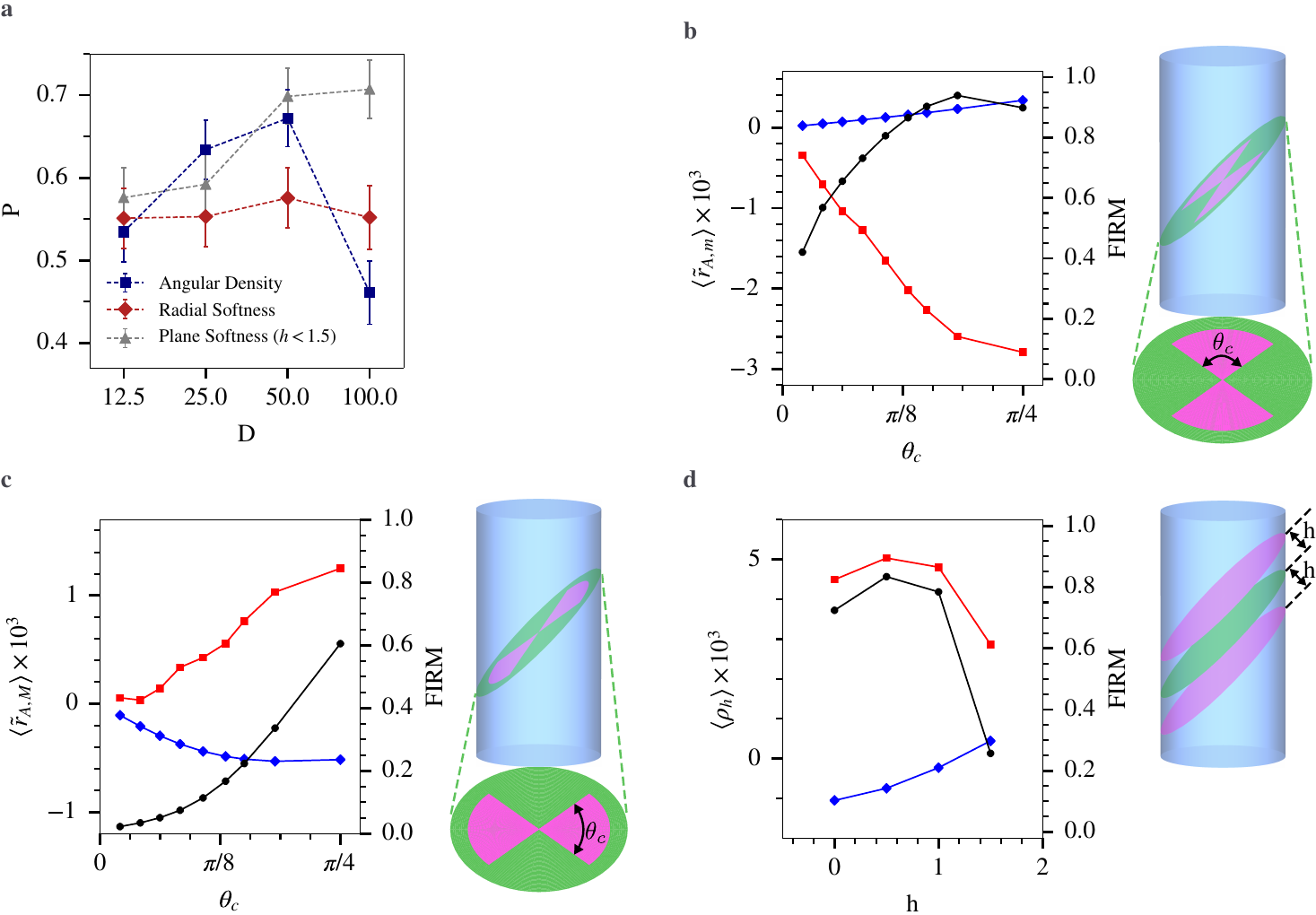}
    \caption{
        \textbf{Fluctuation models for various sets of structure functions.}  
        \textbf{(a)} The test set accuracy of the fluctuation models based on all angular density structure functions, the average softness structure functions as a function of radial position, and the average softness of planes $h \leq 1.5$ from the plane of consideration against all pillar diameters, $D$. Plots of the residuals of the angular density structure functions along the
        \textbf{(b)} minor, $\langle r_{A, m} \left( i; 3.00, 48.6, \theta_{c} \right) \rangle$, and
        \textbf{(c)} major, $\langle r_{A, M} \left( i; 3.00, 48.6, \theta_{c} \right) \rangle$, for the $D = 50$ pillars. 
        \textbf{(d)} Plots of the residuals of the plane softness structure functions ($\langle \rho_{h} \left( i; 0.5, h \right)  \rangle$) for the $D = 100$ pillars. 
        For plots of residuals listed above, the FIRM score corresponds to the given fluctuation model, not the plane weakness measure found using all structure functions. A tilde above the residual function indicates that the residuals have been normalized by the maximum of the corresponding \textit{original} structure function set averaged over all planes.
    }
    \label{fig:FluctuationModels}
\end{figure}

We use test set accuracy as a metric of the predictive strength of various fluctuation models, and the results of this analysis are shown in Figure \ref{fig:FluctuationModels}a. The fluctuation models based on the fluctuations of all of the angular density structure functions do no better than chance ($P = 50\%$) at $D = 12.5$ and $D = 100$ but do exhibit some predictive power at intermediate pillar diameters. To understand the predictive nature of these fluctuatons, we denote these residuals of the minor and major angular density structure functions as $r_{A,m}$ and $r_{A,M}$ respectively. FIRM scores listed describe the percentage of variance in the \textit{fluctuation model} that is described by each residual. Here we see the minor angular structure functions in Figure \ref{fig:FluctuationModels}b are quite undercoordinated and become increasingly more so with larger angular resolution. In contrast, the major angular structure functions in Figure \ref{fig:FluctuationModels}c are overcoordinated compared to the average plane. This suggests that the undercoordination experienced by shear band planes at these intermediate pillar diameters, between $25$ and $50$ particle diameters, typically occurs along its minor axis. As the pillar diameter grows, the size of these fluctuations decrease as a percentage the plane's radius. This leads to a decrease in the importance of these fluctuations at large pillar diameters. In small pillars, shear banding is entirely controlled by density fluctuations in pillar planes rather than the geometry of these fluctuations.
 
Next, we turn to fluctuation models based on the fluctuations of the radial softness structure functions. {\it A priori}, we might expect fluctuation models based on these structure functions to be the most predictive of all the mean softness models due to these structure functions' large MFIRM scores relative to the other softness-based structure functions. Instead, Figure \ref{fig:FluctuationModels}a shows that these models have test set accuracies of just higher than chance, approximately $55$ percent. Because these structure functions have such high MFIRM scores but are not very predictive on their own, these structure functions must be highly correlated with the plane or radial structure functions. Because this effect is not independent of radial fluctuations in the pillar diameter, we presume that much of this effect is due to enhanced surface mobility, which is commonly found in glassy materials with free surfaces \cite{paeng_direct_2011, zhang_long-range_2016, sussman_disconnecting_2017}. Particles near the surface are more mobile, potentially allowing them to explore phase space locally \cite{lyubimov_orientational_2015} and leading to harder structures due to a slower effective quench rate \cite{schoenholz_relationship_2017}. Thus, shear band planes which tend to have smaller local radii are likely to have harder particles at small $R$ than the average plane. Figure \ref{fig:FIRM}c also supports this idea as we find that on both on average and in shear band planes, softness decreases as we approach the surface of the pillar.

Finally, we examine fluctuation models based on the fluctuations of the plane softness averaged over the entire plane. For simplicity of interpretation, we restrict our analysis to the mean softness of planes that are local to the test plane, $h \leq 1.5$. Although the plane softness structure functions have the smallest MFIRM scores out of all of the sets of structure functions we have examined, their fluctuation models obtain large test set accuracies ($P = 0.71 \pm 0.04$) at large pillar diameters. This indicates that they must measure some latent variable not covered by the simple model involving only the plane and radial density; i.e., the specific packing in the shear band plane becomes increasingly important as the pillar diameter increases. To understand this latent variable, we plot the residuals $\rho_h$ of the plane softness structure functions in Figure \ref{fig:FluctuationModels}d for the $D=100$ pillar. Here, we see that shear band planes are softer than the average plane in the pillar ($h = 0$). This effect is apparently important since the FIRM scores suggest that the variance of each of the first three structure functions accounts for approximately $70$ percent of the variance in the fluctuation model. We find that the mean softness of shear band planes decreases sharply at $h = 1.5$, and adding additional plane softness or angular softness structure functions to this model does not improve its accuracy (See supporting information). 

Taken together, our analysis of the fluctuation models suggests that as we approach the large pillar limit, the only latent variable that is predictive of shear banding and not accounted for by the plane's radial fluctuations is the mean softness. This is interesting as the importance of these radial fluctuations is decreasing with increasing pillar diameter as shown by the MFIRM scores of the radial and plane density structure functions in Figure \ref{fig:MFIRM}a. Therefore, we expect softness, a microscopic structural quantity to play a major role in the macroscopic dynamics, and the identification of such a structural quantity is a key step for the development of mesoscale and constitutive models for the dynamics of materials \cite{ottinger_beyond_2005}. 

\section{CONCLUSION}

In summary, our results show that the mesocopic structure of planes can be used to predict shear banding in amorphous solids. This structure can be quantified by plane weakness. According to our analysis, the main component of plane weakness for submicroscopic pillars are small, less than $\frac{1}{2}$ of a particle diameter, radial fluctuations on the exterior of the plane. These fluctuations come from the thermalization of the pillar alone and are not artificially induced. This provides valuable insight about manufacturing strong nanoscale components: to strengthen nanoscale components we may neglect bulk effects and focus on developing components that are smooth on the atomistic level. Even in pristine lab environments, surface defects large enough to cause shear banding may arise in the melt of a material.

As pillar diameter increases, this variable becomes less important and is replaced by other structure functions. In particular, we find that the mean softness local to a plane is an increasingly important predictor of shear banding with increased pillar diameter and is the dominant predictor outside of the radial fluctuations at the largest pillar diameter considered. This observation links the machine learned quantity softness to mesoscale theories such as Shear Transformation Zone (STZ) theory which hypothesize mesocale ``configurational soft spots", regions that are more likely to yield under shear stress \cite{manning_strain_2007}. This link is non-trivial as softness is constructed as a measure of short, local particle motions while shear bands are by definition long timescale, non-local events. Moreover, because we are only using configurational information prior to deformation to predict  shear bands, we have shown that at temperatures well below the glass transition that these defects can be considered to be frozen in place, i.e. we do not need to consider thermal fluctuations to build a mesoscale model that predicts mechanical behavior so long as such behavior occurs well below $T_g$ even when the constituent pieces of a material are atomic in nature.


\section{METHODS}

\section{Simulation Model}
We simulate a coarse grained bead-spring polymer with chains of length $N = 5$. The bonded interactions are taken through a harmonic bonding potential,
\begin{equation}
\label{eq:1}
U_{jk}^{b}=\frac{k_h}{2}\left(r_{jk} - d \right)^2,
\end{equation}
where $r_{jk}$ is the radial distance between monomers $j$ and $k$ and $k_h = 2000 \epsilon / d^2$. Here, $d$ and $\epsilon$ are the length and energy scales of our simulations respectively. The non-bonded interactions are taken using a modified 12-6 Lennard-Jones (LJ) potential,
\begin{equation} 
\label{eq:2}
U_{jk}^{nb}=4 \epsilon \left[ \left( \frac{\sigma}{r_{jk}-\Delta} \right)^{12} - \left( \frac{\sigma}{r_{jk}-\Delta} \right)^{6}  \right].
\end{equation}
We choose $\Delta = 0.75 d$ and $\sigma = d - \Delta / 2^{1/6}$. This gives our potential shorter range and higher curvature while restricting the minimum to reside at the same location as the standard LJ potential where $\Delta = 0$. This modification promotes brittle fracture at low temperatures as is expected in experiments. In the text, we present our findings in units reduced by $d, \epsilon$ and the monomer mass $m$. This study was completed using the LAMMPS~\cite{plimpton_fast_1995} simulation package with a simulation timestep of $0.0006636$. The pillars are aligned along the $\hat{z}$ axis and periodic in this direction, and surfaces in the radial direction are free. We hold the length of our pillars fixed at $L = 200$ particle diameters and vary the diameter of our pillars to be nominally $D = 12.5$, $25$, $50$, and $100$ particle diameters. We generate $N_{\text{pillar}} = 100$ independent pillar configurations for the three smallest pillar diameters and $N_{\text{pillar}} = 50$ independent pillar configurations for the largest diameter pillars.

Using a cooling rate of $5  \times 10^{-5}$, we find the glass transition temperature of the pillars to be $T_g = 0.38$ by identifying the intersection of linear fits of the density as a function of temperature in the supercooled and glassy states. Pillars were thermalized at $T = 0.5$ within a cylindrical, harmonic confining wall which is fixed to ensure the density of the monomers is $\rho = 0.3$. The pillars were cooled at a rate of $5 \times 10^{-4}$ to a temperature of $T = 0.05$. This caused the pillar diameter to contract away from the confining wall as the density of monomers rose to $\rho = 1.0$ below $T_g$.

\subsection{Development of Softness Field}

The softness field used in this study was first characterized in Ref.~\citenum{cubuk_structure-property_2017}. We repeat relevant details here for completeness. This field is developed in a similar way to plane weakness. We first characterize the local structure around each particle $j$, using a set of ``local structure functions":

\begin{equation}
\Psi_{R} (j;\mu, L) = \sum_{k} \mathrm{e}^{(r_{jk}-\mu)^2/L^2}
\label{eqM:1}
\end{equation}

\begin{equation}
\Psi_{A} (j;\xi,\lambda,\zeta) = \sum_{k,l} \mathrm{e} ^{\left(r_{jk}^2+r_{kl}^2+r_{jl}^2\right)/\xi^2}\left(1+\lambda \cos \theta_{jkl} \right)^\zeta
\label{eqM:2}
\end{equation}

where $\mu$, $L$, $\xi$, $\lambda$, and $\zeta$ are parameters that characterize the members of each family of structure functions. Here, $r_{jk}$ is the distance between particles $j$ and $k$. The variable $\theta_{jkl}$ is the angle made between particles $j$, $k$, and $l$. The summations are performed for all particles within a radius $R_c^S$. Our results are insensitive to changes in $R_c^S$ so long as we include the first few neighboring shells \cite{cubuk_identifying_2015}. In this work, we set $R_c^S = 2.5$. The parameter sets that we used to characterize the local environment may be found in the supporting information.

Next we need to develop a training set of rearranging and non-rearranging particles. To create this set, we ran additional independent molecular dynamics simulations in which we thermalized and strained pillars at several temperatures: $T = 0.05$, $0.1$, $0.15$, $0.2$, $0.25$, $0.275$, $0.3$, and $0.325$. These pillars all had a nominal diameter of $D = 50$ and had a length along their $\hat{z}$ axis of $100$. Because the deformation of the pillars causes affine transformations of particle configurations which do not necessarily correspond to rearrangements, we quantify rearrangements of particle $j$ using:

\begin{equation}
D^{2}_{\text{min}} (j;t) = \frac{1}{N_j} \sum_{k}^{N_j} [\boldsymbol{r}_{jk}(t+\Delta t)-\boldsymbol{\Lambda}_{j}(t) \boldsymbol{r}_{jk}(t) ]^{2}
\label{eqM:3}
\end{equation}

which measures the non-affine motion of particle $j$ at time $t$. Here $\boldsymbol{r}_{jk}$ is the vector between particles $j$ and $k$ and $\boldsymbol{\Lambda}_{j}(t)$ is the best fit local gradient tensor about particle $j$ which minimizes the quantity \cite{falk_dynamics_1998}. Summations are performed over all $N_j$ particles within a cutoff radius of $2.5$ particle diameters. We chose $\Delta t$ to correspond to a strain of $0.00166$. We say that a particle $j$ at time $t$ rearranges if $D^{2}_{\text{min}}(j;t) > 0.1$. This value was chosen by using the same method as in Ref.~\citenum{cubuk_identifying_2015}. Additionally, we confine our rearranging and non-rearranging sets of particles to be selected from a region $8$ particle diameters from the center of the pillar and in the elastic regime of strain to avoid rearrangements caused by zero-modes on the surface of the pillar and particles in the shear band respectively. At each temperature, we chose $N_r = 700$ randomly rearranging particles, and $N_n = 700$ non-rearranging particles to be in our training set. We say that a particle is non-rearranging if it has the one of the lowest $N_n$ values of $D^{2}_{\text{min}}$ averaged over a relaxation time \cite{schoenholz_structural_2016}.

We then use a linear support vector machine (SVM) to calculate the hyperplane that best separates points corresponding to rearranging particles from points corresponding to non-rearranging particles. It is not possible to specify a hyperplane that completely separates rearranging particles from non-rearranging ones. Thus, the SVM is designed to penalize particles whose classification is incorrect. This misclassification penalty is controlled by the parameter $C$ where larger $C$ values correspond to fewer incorrect classifications. This parameter was chosen to be $C=0.1$ by k-folds cross validation. We find that more than $93\%$ of rearrangements occur on particles with softness $S>0$ by nested cross validation \cite{cawley_over-fitting_2010}. As with plane weakness, SVM algorithm was implemented using the scikit-learn package \cite{pedregosa_scikit-learn:_2011}. For the purposes of this study, we normalize our softness field to have zero mean and unit variance at each pillar diameter. This leads to an easier interpretation of our softness based results as the number of standard deviations away from $0$.

\subsection{Description of Structure Functions}

Shear bands are expected to form along approximately $45\degree$ planes in the pillars. We partition our pillars into $N_{\text{plane}}=7200$ $45\degree$--planes with $200$ partitions in the $\hat{z}$ axis and $36$ partitions in the $\hat{\theta}$ direction, along the polar angle. We seek to mathematically encode the structure of these planes. To do this, we divise a set of ``structure functions" that describe the local structure of the pillar around each of plane. We define these functions to respect the symmetries of the elliptical prism that characterizes each plane in the pillar. These functions come in two categories with three families each. The first category is the density structure functions:

\begin{equation}
G_{h} \left( i; \xi_{h}, h \right) = \frac{1}{D^{2}} 
\sum_{j} \Theta_{ij}^{P} \left(h,  \xi_{h} \right)
\label{eq:3}
\end{equation}

\begin{equation}
G_{R}  \left( i; \xi_{h}, L_{R}, R \right) =
\frac{1}{R} \sum_{j} 
\Theta_{ij}^{P} \left(0,  \xi_{h} \right)
e^{-d_{ij} \left( R \right)^{2} / L_{R}^{2}}
\label{eq:4}
\end{equation}

\begin{equation}
G_{A, a}  \left( i; \xi_{h}, \xi_{R}, \theta_c \right) = 
\frac{1}{D^{2}} \sum_{j}
\Theta_{ij}^{P} \left( 0, \xi_{h} \right)
\Theta_{ij}^{E} \left( \xi_{R} \right)
\cos \left( \theta_{ij}^{a} \right)^{\zeta \left( \theta_c \right)}
\label{eq:5}
\end{equation}

where each structure function is for a plane $i$ and sums are performed over all particles $j$ whose contribution to the sum is greater than $0.1$ for numerical efficiency. Here, $L_{R}$, $\xi_{h}$, $\xi_{R}$, $h$, and $R$ are parameters that characterize these functions. The function $d_{ij}\left( R \right)$ is the distance in plane $i$ that particle $j$ is away from an ellipse that is centered on the $\hat{z}$ axis and has a minor axis of length $R$. This distance is found numerically using the algorithm in Ref.~\citenum{eberly_distance_2016}. The ellipse is defined by the equation $ (x^{M})^{2} / 2 + (x^{m})^{2} = R^{2}$ where $x^{M}$ and $x^{m}$ are the in plane distances along the major and minor axes respectively. The function $\Theta_{ij}^{E}\left( \xi_{R} \right) = e^{-\left( (x^{M}_{ij})^{2} / 2 + (x^{m}_{ij})^{2} \right) / \xi_{R}^{2}}$ is a soft step function for particles within an ellipse with a minor axis of length $\xi_R$. The variable $\theta^{a}_{ij}$ is the angle between the $a$ axis of plane $i$ and particle $j$ where $a$ is either the major $(M)$ or minor $(m)$ axis. Here, $\zeta \left( \theta_c \right) = \frac{-1}{\log_{2} \left( \cos(\theta_{c}) \right)}$.

These families correspond to simple physical quantities in the following way. Eq~\ref{eq:3} is proportional to the density of particles a distance $h$ away from plane $i$ in a plane of thickness $\xi_{h}$. Eq~\ref{eq:4} is proportional to the density of particles in an elliptical shell of width $L_R$ and thickness $\xi_{h}$ that has a minor axis of length of $R$ and is centered on plane $i$. Finally, $\zeta \left( \theta_{c} \right)$ is defined so that the $\cos \left( \theta_{c} \right)^{\zeta \left( \theta_{c} \right)} = \frac{1}{2}$ allowing us to interpret of this term as another soft step function with a cutoff angle of $\theta_{c}$.  Thus, Eq~\ref{eq:5} is proportional to the density of particles in pie slices that have width $\theta_c$ and width of $\xi_{R}$ and depth of $\xi_{h}$ along the major and minor axes of plane $i$. We call these families of structure functions the plane density, radial density, and angular density structure functions respectively.

The other category is the softness structure functions. These come in three families, $\Gamma_{h} \left( i; \xi_{h}, h \right)$, $\Gamma_{R}  \left( i; \xi_{h}, L_{R}, R \right)$, and $\Gamma_{A, a}  \left( i; \xi_{h}, \xi_{R}, \theta_c \right)$, and measure the mean softness of the regions that correspond to the density structure functions, $G_{h} \left( i; \xi_{h}, h \right)$, $G_{R}  \left( i; \xi_{h}, L_{R}, R \right)$, and $G_{A, a}  \left( i; \xi_{h}, \xi_{R}, \theta_c \right)$ respectively. We define these functions specifically as:

\begin{equation}
\Gamma_{h} \left( i; \xi_{h}, h \right) = \frac{
\sum_{j} S_{j} \Theta_{ij}^{P} \left(h,  \xi_{h} \right)}{
\sum_{j} \Theta_{ij}^{P} \left(h,  \xi_{h} \right)} 
\label{eqM:4}
\end{equation}

\begin{equation}
\Gamma_{R}  \left( i; \xi_{h}, L_{R}, R \right) =
\frac{\sum_{j} S_{j} \Theta_{ij}^{P} \left(0,  \xi_{h} \right)
e^{-d_{ij} \left( R \right)^{2} / L_{R}^{2}}}
{\sum_{j} \Theta_{ij}^{P} \left(0,  \xi_{h} \right)
e^{-d_{ij} \left( R \right)^{2} / L_{R}^{2}}}
\label{eqM:5}
\end{equation}

\begin{equation}
\Gamma_{A, a}  \left( i; \xi_{h}, \xi_{R}, \theta_c \right) = 
\frac{\sum_{j} S_{j}
\Theta_{ij}^{P} \left( 0, \xi_{h} \right)
\Theta_{ij}^{E} \left( \xi_{R} \right)
\cos \left( \theta_{ij}^{a} \right)^{\zeta \left( \theta_c \right)}}{\sum_{j}
\Theta_{ij}^{P} \left( 0, \xi_{h} \right)
\Theta_{ij}^{E} \left( \xi_{R} \right)
\cos \left( \theta_{ij}^{a} \right)^{\zeta \left( \theta_c \right)}}
\label{eqM:6}
\end{equation}

where each function is for a plane $i$ and sums are performed over all interior particles $j$. For this study, we define the interior of the pillar as all particles greater than $3.5$ particle diameters from the pillar's surface. Summations are restricted to interior particles because the structures which cause rearrangements in the bulk, where the softness field was developed, are likely to be different than the structures on the surface of the pillars that lead to rearrangements. For numerical efficiency, we further restrict the summation so that a term only contributes to either sum if the product of that term's functions (excluding $S_{j}$) is greater than $0.1$. We call these structure functions the plane, radial, and angular softness structure functions respectively.

\subsection{Training and Parameter Selection}
For each pillar, we describe every 45\degree--plane prior to deformation with $M=612$ structure functions (See supporting information). At each pillar diameter, we standardize each structure function by subtracting the mean and dividing by the standard deviation. We then assign each plane $i$ a vector, $\boldsymbol{p}_{i} \in \mathbb{R}^{M}$ where each orthogonal component of the vector is one of the standardized structure functions. We call these the ``structure vectors", $\{\boldsymbol{p}_{1}, ..., \boldsymbol{p}_{N}\}$ where $N = N_{\text{plane}} \times N_{\text{pillar}}$.

We now evaluate the local von Mises shear strain rate between the unstretched pillar configuration and the pillar configuration at a strain of $\epsilon = 5.5 \%$ with a cut-off radius of $2.5$ particle diameters. At all pillar diameters, we see strain localization for this strain. For each pillar we evaluate the quantity, 

\begin{equation}
\langle J_{2} \rangle_{j} = 
\frac{\sum_{i} J_{2,i} \Theta_{ij}^{P}\left( 0, \xi_{h} \right)}
{\sum_{i}\Theta_{ij}^{P}\left( 0, \xi_{h} \right)}.
\label{eq:8}
\end{equation}

where $J_{2,i}$ is the local von Mises strain rate of particle $i$ where the summation runs over the interior of the pillar. Here, we take $\xi_{h} = 2$. We pick the planes with the maximum and minimum values of $\langle J_{2} \rangle_{j}$ as shear band and non-shear band planes for each pillar studied. This yields a training set with $2 N_{\text{pillar}}$ elements at each pillar diameter.

Two choices are made in the development of our linear SVM used to generate plane weakness. First, we must decide which features to allow in our linear SVM. We limit the features used in our fit in order to prevent overfitting our model to noise in our data. Second, the SVM method typically incorporates a misclassification penalty $C$, as described in the Development of Softness Field section, which must be chosen as well. We want to make both of these choices so that our model best generalizes to new planes. To find the optimal features, we use recursive feature elimination (RFE) \cite{guyon_gene_2002}. The RFE algorithm starts with the initial $M$ structure functions and prunes the least important structure function at each step. 
 
We use stratified 3-fold cross validation with a grid-search technique on the training set to determine the $C$ value and the step on which to terminate the RFE algorithm, $m$. Here, we select a set of possible $C$ values ranging from $10^{-4}$--$10^{0}$. Then, we partition the training set into $k = 3$ ``folds" with equal numbers of shear band and non-shear band planes in each. We denote $1$ of these folds the ``validation set". For each $C$, we eliminate structure functions using the RFE algorithm on planes from the other $2$ folds until $1$ structure function remains. At each step of the algorithm, we record the percentage of correctly classified planes in the validation set due to a linear SVM developed  using the other $2$ folds. This process is repeated with each of the $3$ folds being used as validation sets. We randomly shuffle the planes between the $3$ folds and repeat this procedure $10$ times to ensure that our parameter selection is independent of how the folds are selected. We say the $m$ and $C$ values which best generalize to new data are those which produce the highest average percentage of correctly classified planes across all folds and re-shufflings. To determine the structure functions used in the final model, we run the RFE algorithm at the most generalize-able $C$ for the $m$ steps on entire training set. We train a final linear SVM with the remaining $M' = M - m$ structure functions at the most generalize-able $C$. To embed this hyperplane into $\mathbb{R}^{M}$, we simply add $0$'s to the plane normal at every location in which a feature was removed.

\subsection{Measures of Binary Classification Performance}

We want to obtain an unbiased estimate of how well our model will generalize to data outside of the training set. We use nested stratified k-folds cross validation to do this \cite{cawley_over-fitting_2010}. We partition our data into $k = 10$ folds with an equal number of shear band and non-shear band planes in each. We retain one of these folds as a ``test set". Then, we perform feature selection and train a linear SVM using the planes from the other folds. For the resulting linear SVM, we measure the percentage of correctly classified planes in the test set. We repeat the process using each of the other folds as test sets. To ensure that our results do not depend on how the folds are chosen, we randomly shuffle the planes between the folds and repeat the process $10$ times. We average our results to obtain the ``test set accuracy" of our classifier. Because the test set accuracy was obtained by using planes that were not used in fitting our classifier, it is a good measure of how well our classifier will generalize to planes outside of the training set. Similarly, we obtain the expected percent shear bands on weak planes, $P(W>0|SB)$, by looking at the average percentage of correctly classified shear band planes in the test sets.

\subsection{Development of MFIRM}

MFIRM is an extension of FIRM \cite{zien_feature_2009} which allows us to calculate the importance of a set of $N$ structure functions. Let 

\begin{equation}
\boldsymbol{f}: \mathbb{R}^M \xrightarrow{} \mathbb{R}^{N}
\label{eqM:7}
\end{equation}

be a function which projects the orthogonal components which correspond to the set of structure functions from the original vector space of all structure functions to a new vector space with only the structure functions of which we wish to find the importance. The expected plane weakness given a set of values of the selected features $\boldsymbol{t} \in \mathbb{R}^N$ is:

\begin{equation}
q_{\boldsymbol{f}} \left( \boldsymbol{t} \right) = \langle W \left(\boldsymbol{p} \right) | \boldsymbol{f} \left( \boldsymbol{p} \right) = \boldsymbol{t} \rangle
\label{eqM:8}
\end{equation}

The MFIRM score of this set of features then corresponds to the standard deviation of $q_{\boldsymbol{f}} \left( \boldsymbol{t} \right)$: 

\begin{equation}
Q_{\boldsymbol{f}} = \sqrt{
\int d \boldsymbol{t}
\left(q_{\boldsymbol{f}} \left( \boldsymbol{t} \right) - \langle q_{\boldsymbol{f}} \rangle \right)^2
P \left( \boldsymbol{f} \left( \boldsymbol{p} \right) = \boldsymbol{t} \right)
}
\label{eqM:9}
\end{equation}

where $P \left( \boldsymbol{f} \left( \boldsymbol{p} \right) = \boldsymbol{t} \right)$ is the probability density of obtaining selecting the structure function values $\boldsymbol{t}$ and $\langle q_{\boldsymbol{f}} \rangle$ is the expected value of $q_{\boldsymbol{f}} \left( \boldsymbol{t} \right)$.

In general, this quantity is quite difficult to calculate as $P \left( \boldsymbol{f} \left( \boldsymbol{p} \right) = \boldsymbol{t} \right)$ is unknown. To simplify calculation, we assume the distribution of structure functions is normally distributed with a mean of $\boldsymbol{\mu}$ and covariance matrix $\boldsymbol{\Sigma}$. The mean may be partitioned into $\boldsymbol{\mu_\text{f}}$ and $\boldsymbol{\mu_\text{l}}$ for which correspond to the sets of structure functions that we wish to know the importance of and leftover structure functions that are not in that set. Similarly, we may partition the covariance matrix as well,

\begin{equation}
\boldsymbol{\Sigma} = 
  \left(\begin{array}{@{}ccc}
    \boldsymbol{\Sigma}_{\text{ll}} & \boldsymbol{\Sigma}_{\text{lf}} \\
    \boldsymbol{\Sigma}_{\text{fl}} & \boldsymbol{\Sigma}_{\text{ff}}
  \end{array}\right).
\label{eqM:10}
\end{equation}

Then, via the properties of the conditional distributions of the multivariate normal distribution, we find

\begin{equation}
q_{\boldsymbol{f}} \left( \boldsymbol{t} \right) - \langle q_{\boldsymbol{f}} \rangle = \boldsymbol{n}_{\text{l}}^{T} \boldsymbol{\Sigma}_{\text{lf}} \boldsymbol{\Sigma}_{\text{ff}}^{-1} \left(\boldsymbol{t}-\boldsymbol{\mu_\text{f}}\right)
+
\boldsymbol{n}_{\text{f}}^{T}
\left(\boldsymbol{t}-\boldsymbol{\mu_\text{f}}\right),
\label{eqM:11}
\end{equation}

where $\boldsymbol{n}_{\text{f}}$ and $\boldsymbol{n}_{\text{l}}$ is the partitioned normal of plane weakness. The superscript $T$'s denote transposition. Then, we may use the quadratic form expectation to show that Eq.~\ref{eqM:9} is

\begin{equation}
Q_{\boldsymbol{f}} = \sqrt{\boldsymbol{v}^{T} \Sigma_{\text{ff}} \boldsymbol{v}},
\label{eqM:12}
\end{equation}

where $\boldsymbol{v}^{T} = \boldsymbol{n}_{\text{l}}^{T} \boldsymbol{\Sigma}_{\text{lf}} \boldsymbol{\Sigma}_{\text{ff}}^{-1}+\boldsymbol{n}_{\text{f}}^{T}$. If the structure functions are not normally distributed, this quantity provides a second-order approximation of MFIRM. Because plane weakness is normalized to have a standard deviation of $1$, $Q_{\boldsymbol{f}}$ may be readily interpreted as the percentage of variance in plane weakness that can be described by a given set of features. For models which are not normalized, we can normalize by the standard deviation in the measure to obtain the same interpretation.

\begin{acknowledgement}

This research was supported by the National Science Foundation through award CMMI-1536914 and the University of Pennsylvania Materials Research Science and Engineering Center (MRSEC) (DMR-1720530), including its computational facilities. Other computational facilities employed were provided by XSEDE through allocation TG-DMR150034.

\end{acknowledgement}

\begin{suppinfo}

The following files are available free of charge.
\begin{itemize}
  \item SupportingInformation.pdf: figures and information referenced in text
\end{itemize}

\end{suppinfo}

\providecommand{\latin}[1]{#1}
\makeatletter
\providecommand{\doi}
  {\begingroup\let\do\@makeother\dospecials
  \catcode`\{=1 \catcode`\}=2 \doi@aux}
\providecommand{\doi@aux}[1]{\endgroup\texttt{#1}}
\makeatother
\providecommand*\mcitethebibliography{\thebibliography}
\csname @ifundefined\endcsname{endmcitethebibliography}
  {\let\endmcitethebibliography\endthebibliography}{}

\end{document}